# Moiré circuits: engineering magic-angle behaviors

Weixuan Zhang[1*], Deyuan Zou[1*], Qingsong Pei[1], Wenjing He[2], Houjun Sun[2$], and Xiangdong Zhang[1+]

[1]Key Laboratory of advanced optoelectronic quantum architecture and measurements of Ministry of Education, Beijing Key Laboratory of Nanophotonics & Ultrafine Optoelectronic Systems, School of Physics, Beijing Institute of Technology, 100081, Beijing, China

[2] Beijing Key Laboratory of Millimeter wave and Terahertz Techniques, School of Information and Electronics, Beijing Institute of Technology, Beijing 100081, China

*These authors contributed equally to this work. [+$]Author to whom any correspondence should be addressed. E-mail: zhangxd@bit.edu.cn; sunhoujun@bit.edu.cn

**Moiré superlattices in the twisted bilayer graphene provide an unprecedented platform to investigate a wide range of exotic quantum phenomena. Recently, the twist degree of freedom has been introduced into various classical wave systems, giving rise to new ideas for the wave control. The question is whether twistronics and moiré physics can be extended to electronics with potential applications in the twist-enabled signal processing. Here, we demonstrate both in theory and experiment that lots of fascinating moiré physics can be engineered using electric circuits with extremely high degrees of freedom. By suitably designing the interlayer coupling and biasing of one sublattice for the twisted bilayer circuit, the low-energy flat bands with large bandgaps away from other states can be realized at various twist angles. Based on the moiré circuit with a fixed twist angle, we experimentally demonstrate the effect of band narrowing as well as the localization of electric energy when a magic value of the interlayer coupling is applied. Furthermore, the topological edge states, which originate from the moiré potential induced pseudomagnetic field, are also observed for the first time. Our findings suggest a flexible platform to study twistronics beyond natural materials and other classical wave systems, and may have potential applications in the field of intergraded circuit design.**

Twisted bilayer graphene has emerged as a promising platform to engineer moiré flat bands near the Fermi energy [1-4]. Reducing bandwidths below long-range Coulomb interactions, various exotic correlation effects could be induced in moiré superlattices [5-28]. These phenomena spark the exploration of correlated electron states in moiré systems [21-28]. Except for correlated many-body phases, recent investigations have shown that magic-angle flat bands also possess non-trivial topological properties at the single-particle level [29]. Because of the extreme sensitivity of moiré bands to the twist angle, the advanced technology is required to fabricate the precisely controlled twisted bilayer graphene. To release the accuracy of the required twist angle, some methods are proposed to tune the interlayer coupling of the moiré superlattice [7, 23-25, 30]. While, due to intrinsic limitations of Van der Waals heterostructures, such as the limited interlayer coupling strength and structural inhomogeneity of graphene sheets [7, 30], the adjustable flexibility is still limited.

Recently, incorporating the twist degree of freedom into classical wave systems have showcased interesting features in manipulating wave dynamics [31-37]. While, due to the difficulty in the sample fabrication and measurement, experimentally engineering magic-angle phenomena in classical wave systems has not been explored up to now. Recent investigations have shown that electric circuits can be used as an extremely flexible platform to investigate many novel quantum phases [38-51]. It is straightforward to ask whether moiré physics can be extended to the field of electric circuits to explore novel moiré phenomena.

In this work, we theoretically construct moiré circuits, and experimentally demonstrate that they can be used as manageable platforms to investigate magic-angle behaviors that are hard to be realized in other systems, including chiral symmetric moiré flat bands and moiré potential induced topological edge states. Our proposal provides a useful laboratory tool to investigate twistronics, and may possess applications for the twist-enabled signal control.

We consider the honeycomb bilayer circuit with nodes being classified into A/B-type sub-nodes, which are analogies to A/B-type sublattices in graphene. A capacitor $C_{\text{intra}}$ is used to connect adjacent circuit nodes in the same layer to form the intralayer coupling. By linking aligned nodes in the bilayer circuit without twisting, the circuit-analog of usual bilayer graphene is achieved. If the bilayer circuit is rotated with an angle $\theta$ with respect to a common node, where $\theta$ is determined by the formula of $\cos(\theta) = (m^2 + n^2 + 4mn)/(2m^2 + 2n^2 + 2mn)$ with $m$ and $n$ being two integers [3], the circuit-analog of twisted bilayer graphene is formed. The top chart in Fig. 1a presents the moiré superlattice with $m$=6 and $n$=7 ($\theta$=5.09°). Red and green shaded regions cover areas with circuit nodes being AA/BB-stacked and AB/BA-stacked, respectively. The pink parallelogram marks the moiré unit cell.

To clearly illustrate the interlayer coupling, we plot the connecting pattern in the bottom chart of Fig. 1a, where the total of 45 pairs of inter-connected nodes are applied. The orange, blue and yellow lines correspond to cases with coupling capacitors being $2C_{\text{inter}}$, $C_{\text{inter}}$ and $0.5C_{\text{inter}}$. In Fig. 1b, we plot the enlarged view for a part of twisted bilayer circuit marked by red stars in Fig. 1a. Letters 'A/B' represent A/B-type circuit nodes. It is shown that A-type nodes in the top layer are linked to B-type nodes of the bottom layer through different capacitors. To realize the effective on-site potential, the inductor and capacitors are selected for grounding on different nodes. Right charts in Fig. 1b show the ground setting of circuit nodes enclosed by frames with consistent colors.

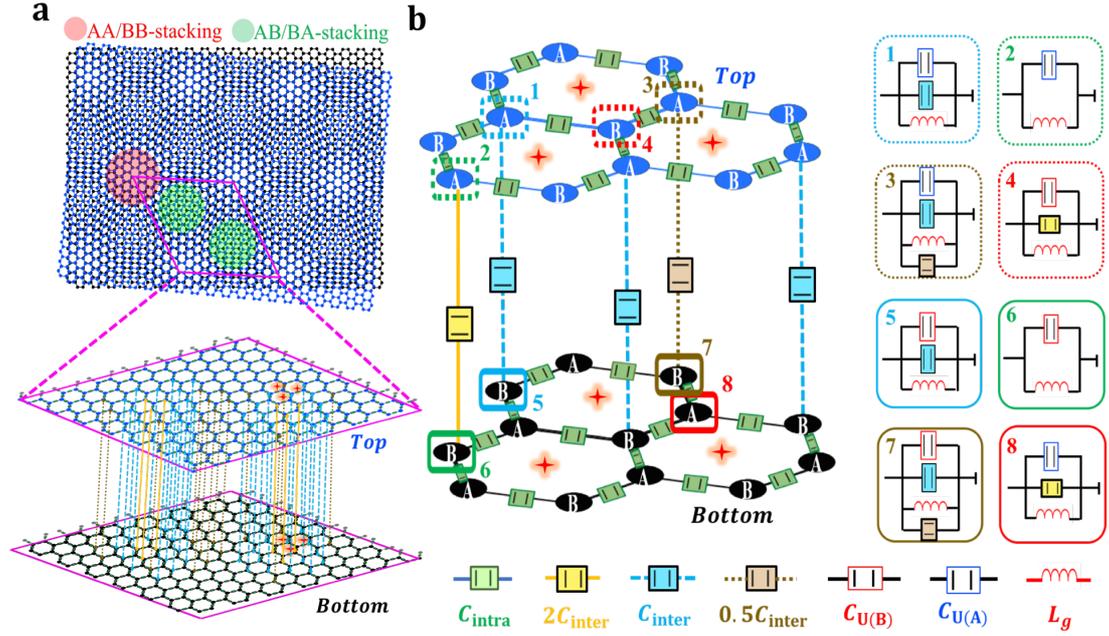

FIG. 1. (a) The moiré superlattice with a commensurate rotation angle being $\theta = 5.09°$ ($m = 6, n = 7$). The pink parallelogram marks the moiré unit cell. (b) The enlarged view of moiré circuits around red stars in Fig. 1a. Eight insets present the grounding of different circuit nodes.

Actually, there are many ways for realizing interlayer couplings of twisted bilayer circuits. Here, we limit the interlayer connection to sustain the chiral symmetry, where only AB/BA-type interlayer couplings are turned on in the region of AB/BA-stacking. This is due to the fact that the chiral symmetric interlayer coupling has been pointed out to be the key for the formation of perfect moiré flat bands [52]. However, such requirement is hard to be realized in graphene or other systems [1-4, 35-37]. With the remarkable advantage for achieving complex node connections, electric circuits can act as an ideal platform to fulfill the chiral symmetric interlayer coupling.

Not only the consistence of node connections, the fascinating moiré physics could also be simulated. Section S1 gives a detailed derivation for identifying the effective tight-binding parameters in terms of circuit elements [53]. In this case, the eigen-energy of moiré superlattice is directly related to the eigen-frequency of twisted bilayer circuit as $\varepsilon = f_0^2/f^2 - 1$ with

$f_0 = 1/2\pi\sqrt{(3C_{\text{intra}} + 2C_{\text{inter}})L_g}$ being equivalent to the zero-energy. Owing to the exact correspondence, twisted bilayer circuits could show various properties of moiré superlattices as demonstrated below. Hence, we called twisted bilayer circuits as 'moiré circuits'.

Band structures of periodic moiré circuits are shown in Fig. 2a, where $C_{\text{intra}}$ ($L_g$) is taken as 1nF (1uH) (same values are used below) and the effective onsite potential is set as zero. Three subplots correspond to circuits with $C_{\text{inter}}$=0.2$C_{\text{intra}}$, $C_{\text{intra}}$ and 2$C_{\text{intra}}$. Brown dash lines mark the position of $f_0$. It is shown that the dispersion of moiré circuits with a fixed twist angle is strongly dependent on the interlayer coupling. Four effective low-energy flat bands appear when a magic value of the interlayer coupling strength $C_{\text{inter}}$=$C_{\text{intra}}$ is achieved. The relationship between the width of lowest circuit band (nearest to $f_0$) and the ratio of $C_{\text{inter}}$/$C_{\text{intra}}$ is shown in the top chart of Fig. 2b (red line). In addition, the change of low-energy bandgap as a function of $C_{\text{inter}}$/$C_{\text{intra}}$ is plotted in the bottom chart (red line). It is shown that the smallest bandwidth appears at a magic value of $C_{\text{inter}}$=$C_{\text{intra}}$. At the same time, the maximal bandgap is produced. The appearance of isolated flat bands with a maximal bandgap away from other high-energy modes is consistent with the feature of twisted bilayer graphene sustaining the chiral symmetry [52].

The enlarged dispersion of low-energy states with $C_{\text{inter}}$=$C_{\text{intra}}$ is presented in Fig. 2c with red lines. It is found that four low-energy moiré bands still possess finite bandwidths. One advantage of circuit lattices is that the effective onsite potential of the A (B) sub-node can be adjusted to further flatten moiré bands. The blue line in Fig. 2c displays the dispersion with $C_{\text{U(A)}}$=0.2$C_{\text{intra}}$ (without $C_{\text{U(B)}}$). With such a biasing of one sub-node, moiré flat bands appear at 2.207MHz and 2.2509MHz with reduced bandwidths. Moreover, we also calculate the variation of low-energy bandwidth and bandgap as functions of $C_{inter}/C_{intra}$, as shown by blue lines in Fig. 2b. It is shown that the magic value of $C_{inter}/C_{intra}$ keeps the same. It is worthy to note that the construction method of moiré circuits is universal for different commensurate rotation angles and interlayer coupling patterns. See detailed discussions in Section S3 [53].

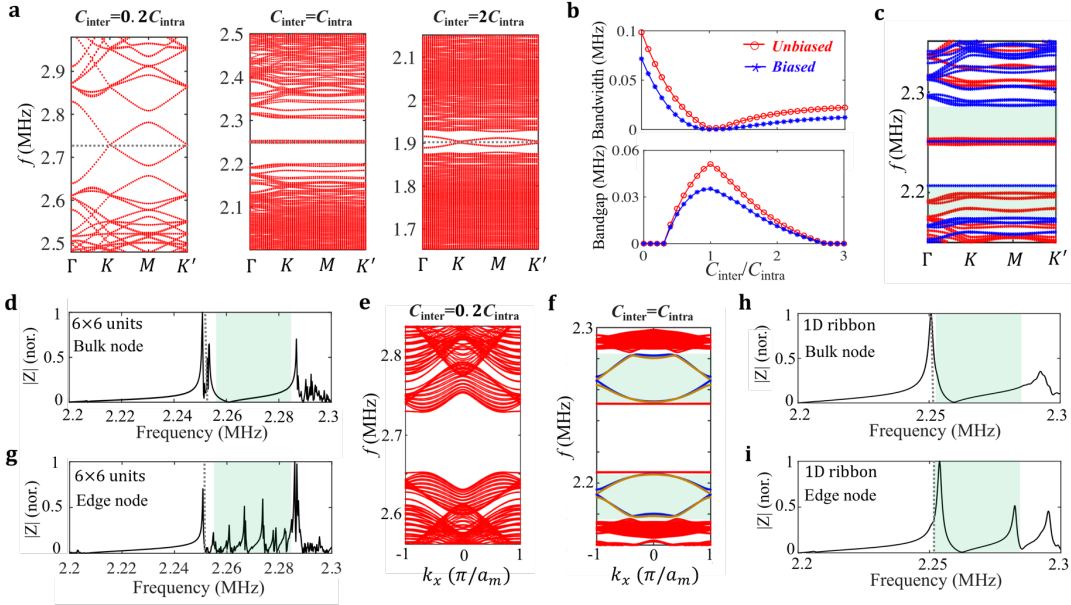

FIG. 2. (a) Band structures of moiré circuits with $C_{inter}=0.2C_{intra}$, $C_{intra}$ and $2C_{intra}$. (b) The relationship between the low-energy bandwidth (and bandgap) and the ratio of $C_{inter}/C_{intra}$ of the biased and unbiased moiré circuits. (c) The blue (red) line presents the enlarged view of band dispersion for the biased (unbiased) circuit with $C_{inter}=C_{intra}$. (d) and (g) present numerical results of the normalized bulk and edge impedances of the biased $6\times6$ moiré circuit with $C_{inter}/C_{intra}=1$. (e) and (f) The dispersion of the biased 1D moiré-ribbon circuit with $C_{inter}/C_{intra}=0.2$ and 1. (h) and (i) The simulated impedances of a bulk and a top edge node in the biased 1D moiré-ribbon circuit with $C_{inter}/C_{intra}=1$.

Then, we perform steady-state simulations of biased moiré circuits with $6\times6$ units. Fig. 2d displays the simulated impedance of a bulk node with respect to the ground with $C_{inter}/C_{intra}=1$. The magnitude is normalized by the maximum. It is shown that impedance peaks appear in a narrow frequency range around $f_0$ (dash line), indicating the excitation of flat bands.

Moreover, the low-energy bandgap (the green region) is also clearly illustrated with the disappear of bulk impedance peaks (other bulk nodes have the same response, see Section S2 [53]).

Except for the flat-band effect, it has been pointed out that low-energy moiré bands also possess non-trivial topological properties [29]. However, the experimental observation of topological edge states is still lacking as far as we know. In the following, we will demonstrate that such non-trivial edge state can be observed in moiré circuits.

A ribbon of moiré circuit, which possesses the translation symmetry along *x*-axis (one unit), and has a finite width (eleven moiré units) along *y*-axis, is considered. Figs. 2e and 2f display dispersion curves of biased moiré-ribbon circuits with $C_{\text{inter}}/C_{\text{intra}}$ being 0.2 and 1. We find that low-energy bulk bands become extremely flat with $C_{\text{inter}}/C_{\text{intra}}$ =1. This is consistent with the dispersion of periodic moiré circuits. More importantly, two pairs of helical edge states appear in non-trivial bandgaps (green regions) with blue (orange) lines corresponding to edge states at the top (bottom) layer. The formation of nontrivial flat bands of the biased twisted bilayer circuit can be explained by the interplay between moiré potential induced pseudomagnetic fields with opposite signs and the massive Dirac point locating at the valley of each monolayer. In this case, two low-energy flat bands originated from the valley of each monolayer can be viewed as two zeroth pseudo Landau levels of Dirac fermions under opposite magnetic fields. Due to opposite signs of pseudomagnetic fields, two moiré flat bands of each valley could carry opposite valley Chern numbers ±1 (the total Chern number is zero due to the existence of time-reversal symmetry), leading to the appearance of helical edge states in nontrivial bandgaps. The influence of twist angle and interlayer coupling on the formation of helical edge states is discussed in Section S4 [53]. It is worthy to note that the biased bilayer circuit possesses unbalanced effective onsite potentials, which can introduce identical mass terms to two valleys of each monolayer and open a trivial bandgap between two pairs of flat bands.

To detect edge states, we simulate the impedance response of an edge node in the moiré circuit with 6 × 6 units, which has lots of edge-localized eigen-modes in the nontrivial bandgap (see Section S5 for details [53]). As plotted in Fig. 2g, it is shown that many impedance peaks appear in the non-trivial bandgap, manifesting the excitation of these in-gap edge states. Moreover, impedance responses of a bulk node and a top edge node in the 1D moiré-ribbon circuit are also calculated, as shown in Figs. 2(h) and 2(i). Since there is one unit along $x$-axis, the band diagram of Fig. 2(f) at $k_x=0$ can be probed. We can see that an impedance peak of the bulk node locates at $f_0$, indicating the existence of the flat band. And, two impedance peaks of the top edge node appear in the non-trivial bandgap with frequencies being consistent with eigen-values of a pair of helical edge states in the top layer at $k_x=0$.

To experimentally observe the moiré-potential induced chiral flat bands and topological edge states, we fabricate the designed moiré circuit. The image is presented at the top chart of Fig. 3a, and the corresponding enlarged view is shown in the bottom chart. Details of the sample fabrication and experimental measurement are provided in Section S6 [53]. Fig. 3b presents the measured impedance spectrum of a bulk node. We find that the impedance peak appears in a narrower frequency range around $f_0$, manifesting the extreme flatten of moiré bands. And, the non-trivial low-energy bandgap can also be identified (green regions), where the impedance of bulk node is nearly vanished. Furthermore, we also test the existence of in-gap topological edge states by measuring the impedance of an edge node, as shown in Fig. 3c. It is seen that impedance peaks of the edge node appear in the effective low-energy bandgap, indicating the excitation of in-gap topological edge states. The lower number of peaks compared to simulations is due to the large loss in experiments, which could make various weak resonances merge together to form lower and smoother peaks. In addition, the disorder effect can induce many extra mini-peaks in bandgaps compared to simulations. Except for above measured circuit nodes, other bulk and edge nodes also possess the same response (see section S2 [53]).

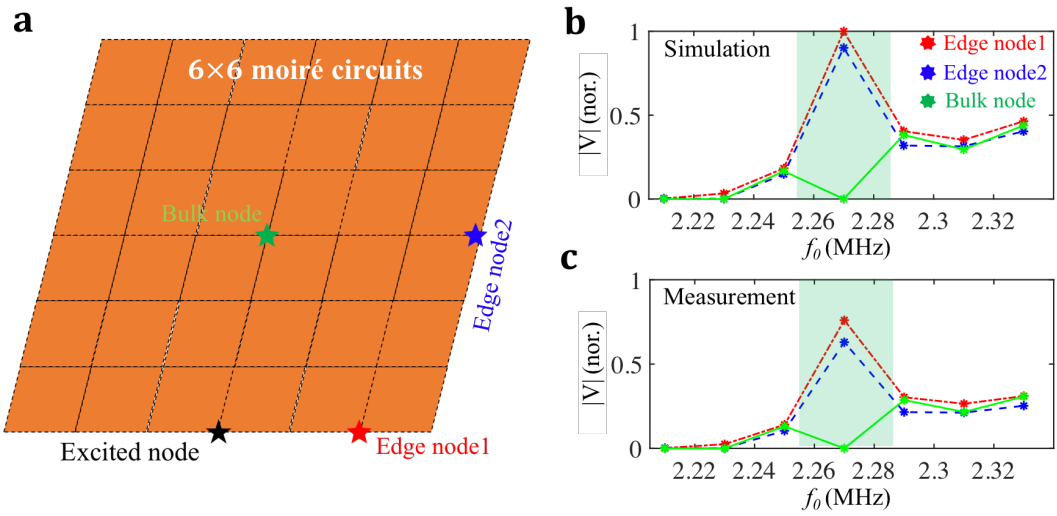

FIG. 3. (a) The top chart displays the photograph image of the moiré circuit ($\theta = 5.09°$) with 6×6 units, and the enlarge view is presented in the bottom chart. (b) and (c) The measured impedances of a bulk node and an edge node in the frequency domain with $C_{\text{inter}}/C_{\text{intra}}=1$. Green regions mark non-trivial bandgaps.

Moreover, moiré circuits could also incorporate the twist degree of freedom to control the voltage signal in a novel way. Here, we demonstrate the twist-enabled energy localization based on the moiré-ribbon circuit. The image of fabricated sample is shown in Fig. 4a. Two boundaries perpendicular to the $x$-axis are connected by $C_{\text{intra}}$ to realize the periodic boundary condition. The measured impedance spectra of a bulk node and a top edge node are shown in

Figs. 4(b) and 4(c). The flat band is illustrated by an impedance peak of the bulk node at $f_0$. Two impedance peaks of the top edge node in the non-trivial bandgap correspond to the excitation of edge states with frequencies being consistent with the band diagram at $k_x=0$ in Fig. 2(f).

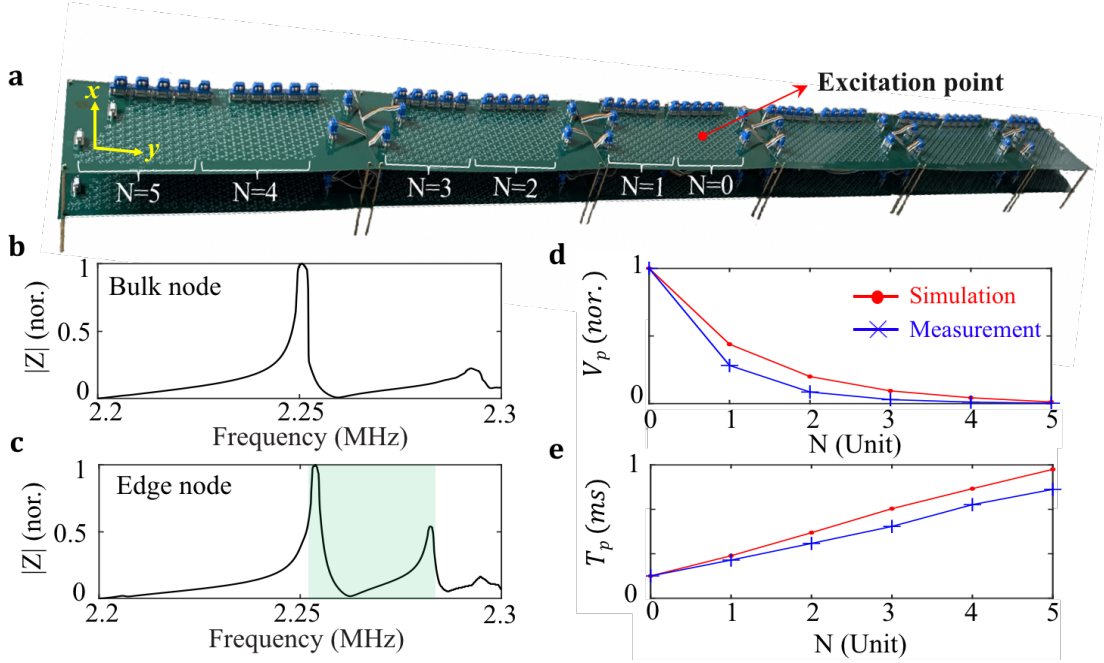

FIG. 4. (a) The photograph image of the fabricated 1D moiré-ribbon circuit. (b) and (c) The measured impedance spectra of a bulk node and a top edge node. (d) and (e) The red and blue lines show the simulated and measured peak value and reaching time of the packet on different circuit nodes.

Then, we inject a voltage packet (with central frequency being $f_0$=2.25MHz) into a bulk node. Red and blue lines in Fig. 4(d) present the simulated and measured results for the normalized peak value of the packet arrived to different circuit nodes (in the unit of a period). Fig. 4e shows the reaching time of packet toward these selected nodes. We find that the packet is exponentially decayed from the excitation node, indicating the localization of the input signal. Moreover, the diffusion speed of the packet is much smaller than that in the circuit without magic interlayer couplings (See Section S7 [53]), manifesting the slow-wave effect induced by moiré flat bands. Such twist-enabled localization and delocalization of electric energy could act as a switch, and the slow-wave effect assisted by moiré flat bands can enhance the circuit nonlinear response.

In addition, the high-efficient transport of voltage packets could be realized based on topological edge states in moiré circuits. The black star in Fig. 5a labels the excited node. Other stars mark detected nodes. Fig. 5b displays simulated peak values of the packet arrived to these circuit nodes with different central frequencies $f_0$. We find that the large (small) peak appears on the detected edge (bulk) node with $f_0$=2.27MHz, which locates in the non-trivial bandgap sustaining edge states. This phenomenon verifies the high-efficient propagation of electronic signal assisted by topological edge states. Fig. S5c presents the measured results, which are consistent with simulations.

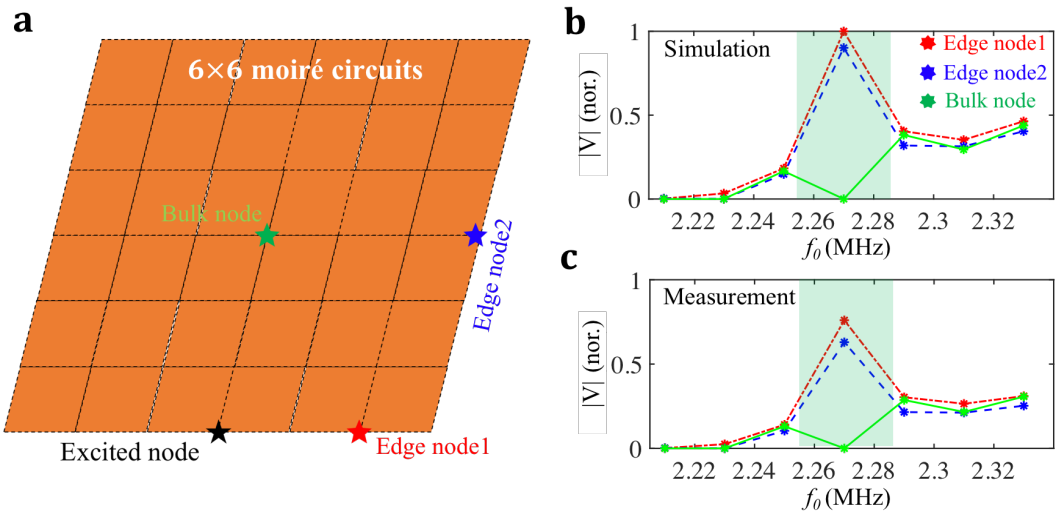

FIG. 5. (a) The schematic diagram of the high-efficient transport of electronic signals assisted by topological edge states. (b) and (c) The simulated and measured peak values of the packet arrived to edge and bulk nodes.

On the other hand, it is worthy to note that low-energy flat bands of moiré materials play a vital role in the formation of strongly correlated phases. Interestingly, electric circuits may also be used as classical platforms to simulate few-body correlated phases in moiré superlattices. A recent work has pointed out that, by mapping the lower-dimensional few-body Hilbert space onto the circuit lattice with higher dimensionality [54], the behavior of strongly correlated bosons described by the two-body Bose-Hubbard model can be simulated with electric circuits. Such a mathematically rigorous mapping can be extended to moiré superlattices with few bosons. In this case, we can experimentally investigate the interplay between particle interactions and flatness of moiré bands, where the interaction-induced phase transition may be observed.

In conclusion, we demonstrate both in theory and experiment that moiré circuits not only provide a flexible platform for observing many fascinating moiré phenomena that are hard to be realized in other systems, but also incorporate the twist degree of freedom into the design of electric circuits for controlling the signal in a novel way, where the sharp localization of electronic energy induced by moiré flat bands and the high-efficient transport assisted by topological edge states are demonstrated as illustrations. Our proposal provides a useful laboratory tool to investigate and visualize many interesting effects related to the twistronics, and may have potential applications in the field of novel circuit design.

**Acknowledgements**

This work was supported by the National key R & D Program of China under Grant No. 2017YFA0303800 and the National Natural Science Foundation of China (No.91850205 and No.61421001).